# Diagnostic Quality Assessment of Fundus Photographs: Hierarchical Deep Learning with Clinically Significant Explanations

Shanmukh Reddy Manne[a], José-Alain Sahel[b], Jay Chhablani[b], Kiran Kumar Vupparaboina[b,*] and Soumya Jana[a]

[a]*Indian Institute of Technology Hyderabad, India*
[b]*University of Pittsburgh School of Medicine, Pittsburgh, USA*

ABSTRACT

Fundus photography (FP) remains the primary imaging modality in screening various retinal diseases including age-related macular degeneration, diabetic retinopathy and glaucoma. FP allows the clinician to examine the ocular fundus structures such as the macula, the optic disc (OD) and retinal vessels, whose visibility and clarity in an FP image remain central to ensuring diagnostic accuracy, and hence determine the diagnostic quality (DQ). Images with low DQ, resulting from eye movement, improper illumination and other possible causes, should obviously be recaptured. However, the technician, often unfamiliar with DQ criteria, initiates recapture only based on expert feedback. The process potentially engages the imaging device multiple times for single subject, and wastes the time and effort of the ophthalmologist, the technician and the subject. The burden could be prohibitive in case of teleophthalmology, where obtaining feedback from the remote expert entails additional communication cost and delay. Accordingly, a strong need for automated diagnostic quality assessment (DQA) has been felt, where an image is immediately assigned a DQ category. In response, motivated by the notional continuum of DQ, we propose a hierarchical deep learning (DL) architecture to distinguish between good, usable and unusable categories. On the public EyeQ dataset, we achieve an accuracy of 89.44%, improving upon existing methods. In addition, using gradient based class activation map (Grad-CAM), we generate a visual explanation which agrees with the expert intuition. Future FP cameras equipped with the proposed DQA algorithm will potentially improve the efficacy of the teleophthalmology as well as the traditional system.

## 1. Introduction

About 90% of global cases of vision impairments are reported in the developing world [30]. Among those, the rural, marginalized and relatively resource-poor populations remain at further higher risk owing to the steep economic and physical barriers to accessing eyecare as well as the general scarcity of local ophthalmologists. A major proportion of the cases are either avoidable or curable with timely detection, diagnosis and treatment [30, 31], and we seek to mitigate those even in low-resource settings. Specifically, we consider a teleophthalmology scenario, where primary eyecare centers serving rural populations are equipped with suitable imaging devices, and clinical investigations are remotely performed based on medical images that are locally acquired and transmitted [35]. However, a significant bottleneck arises because the local technicians are often trained only to operate the imaging device, but not necessarily to differentiate between usable and unusable images in the clinical sense. Thus, in remote clinical investigations, the operator traditionally transmits acquired images, a significant fraction of which may have low diagnostic quality (DQ), and the recipient ophthalmologist assesses several images and repeatedly requests re-acquisition of images till sufficient quantity of usable ones are collected. In such a potentially long-drawn process, the imaging device may need to be engaged multiple times for a single subject, and the time and effort of the ophthalmologist, the operator and the subject are wasted. To make the process more efficient, we propose a frame-work, where an automated algorithm would accurately assess the DQ of the images at the time of acquisition, and instantly prompt the operator to re-acquire additional images, when necessary. The advantage over the traditional framework is schematically depicted in Fig. 1.In this paper, we report the development of the desired diagnostic quality assessment (DQA) algorithm for the ubiquitous imaging modality of fundus photography (FP) that facilitates screening and diagnosis of an array of ocular diseases [1].

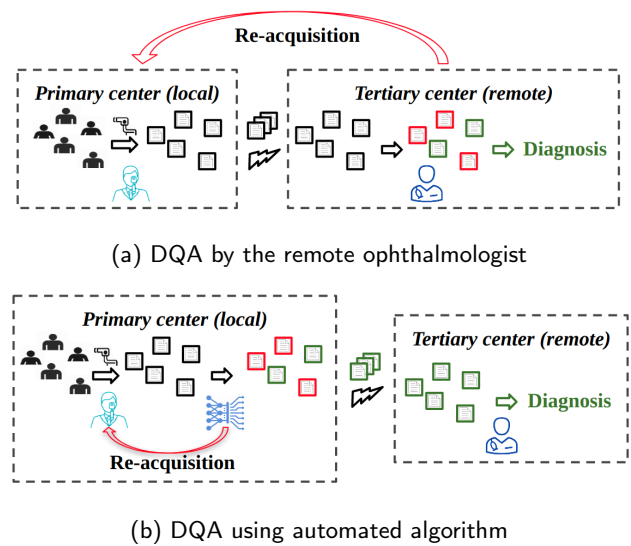

**Figure 1:** Block schematic of teleophthalmological diagnosis: (a) Traditional framework, and (b) Proposed framework enabled by DQA algorithm. (Notation – black avatars: subjects; cyan avatar: technician; blue avatar: ophthalmologist; blue network: automated DQA algorithm (reducing the burden of the ophthalmologist); rectangles with black border: acquired images; with red border: unusable images; with green border: usable images)

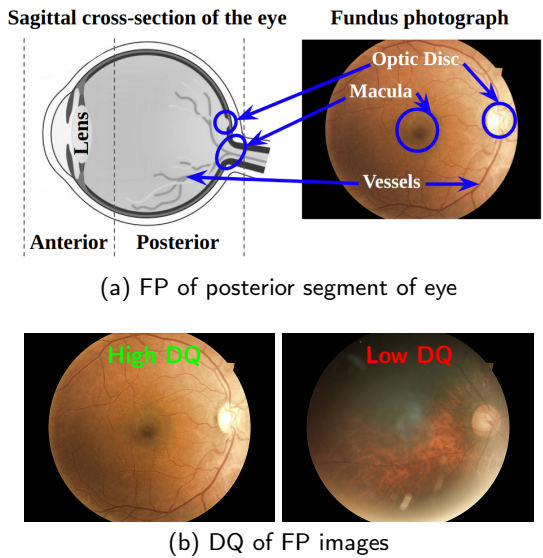

(a) FP of posterior segment of eye

(b) DQ of FP images

**Figure 2:** (a) Sagittal cross-section of eye with corresponding ocular structures (optic disc, macula and vessels, and (b) illustrative examples of FP images with low- and high-diagnostic quality.

FP is a noninvasive procedure, where the posterior portion of eye is photographed, and which allows one to examine the ocular fundus structures such as the macula, the optic disc (OD) and retinal vessels (see Fig. 2 (a)) [9, 41, 24, 28]. Specifically, as part of the preliminary screening of certain prevalent diseases like diabetic reinopathy (DR) and age-related macular degeneration (AMD), the vitreo-retina specialist inspects FP images for possible presence of hemorrhages and geographic atrophies, respectively [18, 46, 26, 2, 27]. When the target disease is glaucoma, the cup-to-disc ratio and peripapillary atrophy are measured to quantify its severity [51, 38, 54]. Evidently, accurate disease screening demands high quality of acquired FP images. However, in practice, low-quality FP images may result from factors including saccadic eye movements, dust or eyelashes on the camera lens, improper illumination or camera focus, and other external conditions [9, 22]. The right image in Fig. 2 (b) depicts such a low-quality image, which resulted from uneven illumination and contrast, and which fails to clearly show ocular fundus structures such as the macula and retinal vessels. In contrast, the left image represents a high quality FP image with the fundus structures clearly visible.

Recent technological advances have significantly reduced the size, the complexity and the cost of FP cameras [33], allowing their deployment to primary centers and train-ing of rural operators with limited technical background, and hence enabling teleophthalmology. Unfortunately, such local operators often have limited knowledge and training to detect quality cues in an acquired FP image, appreciate their clinical relevance, and make decisions regarding re-acquisition. Traditionally, all requests for re-acquisition thus comes from the remote ophthalmologists in the framework of Fig. 1(a), making the teleophthalmology system inefficient. Desiring efficient operation, we propose to equip fundus cameras with a resident DQA tool that automatically accepts only usable images, and rejects low-quality images prompting re-acquisition (Fig. 1(b)).

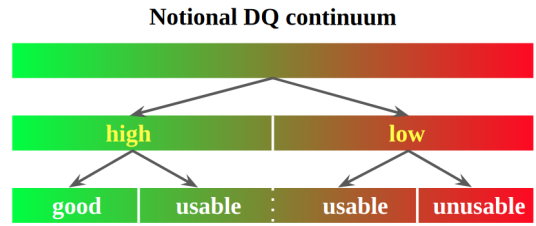

**Figure 3:** Notional DQ continuum and hierarchical assignment of quality labels good, usable and unusable.

The problem of DQA has received markedly less attention compared to that of (no-reference) natural image quality assessment (NIQA) [25]. In contrast to NIQA, where one seeks to quantify the statistical difference of the presented image from the norm without regards to specific spatial features [32, 36, 49], DQA of FP images deals with clarity and visibility of specific fundus structures and ideally requires fresh investigations [9]. Yet, early attempts at distinguishing between high- and low-quality FP images were based on NIQA-inspired features, including pixel histogram [21], image texture, blur matrix [50] [6], distortions in color, contrast, and so on [17], as well as those related to human visual system [47]. Unsurprisingly, the aforementioned methods tended to distinguish FP images in terms of generic quality cues including overall inconsistency in contrast, illumination, focus, and such like, without particular regard to discernibility of fundus structures. In a subsequent FP-specific attempt at DQA, only images centered at macula were considered, and features based on vessel density and 5-bin histogram of each color channel were suggested [12]. More general view-independent techniques have made use of features arising from (i) retinal blood vessel segmentation along with suitable vesselness scoring [19] [34], (ii) structure preserving scattering networks (ScatNet) for grey-scale images [43, 8], and (iii) union of NIQA-inspired and ScatNet features in multi-color space [23], and proven increasingly efficient. Current state-of-the-art methods take legacy deep architectures, pretrained on non-medical images, and minimally train those using labeled FP images to extract fundus-specific deep features [52, 53]. The proposed approach is similar in spirit, but takes a hierarchical approach, which, as explained next, exploits the nature of the problem at hand.

Historically, experts began by grading the quality of FP

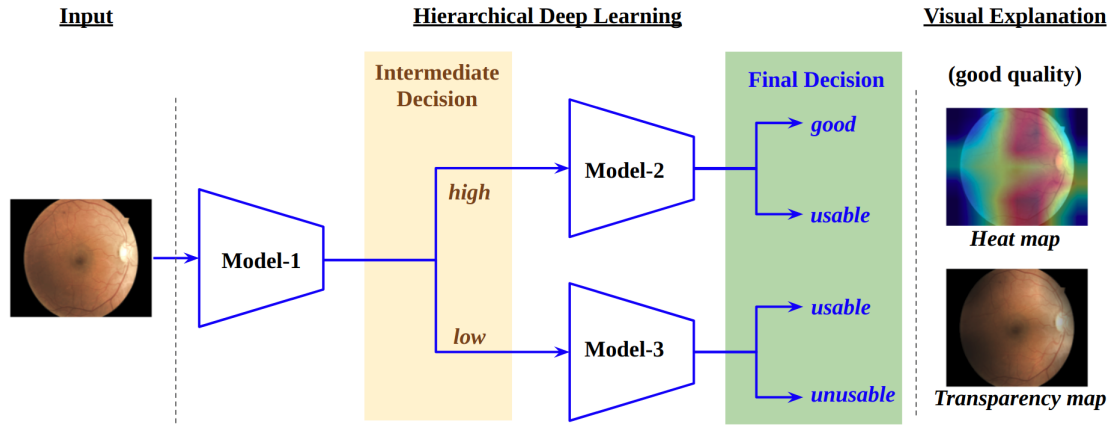

**Figure 4:** DQA of FP images: Proposed workflow.

images as either high or low. Based on FP datasets with such quality labels, binary quality classifiers were developed [42]. However, in practice, while images of medically usable quality may suffice for disease screening, only images exceeding a higher quality threshold are useful for quantification of structures and abnormalities required for proper diagnosis. In one work, to cater to the above requirement based on binary-labeled datasets, the usual two-class classifier was followed by an unsupervised step, where images predicted as having high quality are subdivided into those with truly high quality and those with medically usable quality [4]. However, the accuracy of the follow-up step was not studied, and the possibility of finding usable images among predicted low-quality ones was ignored. To obviate such difficulties, Huazhu *et al.* suggested a natural ternary grading scheme with quality categories, good, usable and unusable ('reject'), and published the publicly available Eye-Quality (EyeQ) dataset with 28,792 annotated FP images [11]. Further, these authors reported a deep learning (DL) architecture that combines individual legacy models for different color spaces, and integrates decision making via one overarching fully connected network. Transfer learning of legacy deep architectures have since been expanded to include statistical features, and spawned a significant number of DQA techniques, which have achieved high performance on the EyeQ dataset [11, 37, 48]. While the state-of-the-art performance is impressive, we hypothesize that there is further room for improvement.

Specifically, viewing the DQ as a notional continuum as depicted in Fig. 3, a traditional three-class classifier needs to simultaneously learn two decision boundaries marking the unusable-to-usable and the usable-to-good transitions. Compare this to a binary classifier, where notions of the extremes of high and low quality are fixed and well separated, and a decision boundary separating those can be learnt somewhat straightforwardly. In contrast, the present three-class classifier deals not only with good and unusable quality (whose extremes are, respectively, same as those of high and low quality), but also with the usable category which is intermediate and does not admit a notion of extreme. Compared to directly learning the boundaries of the usable category, a method of learning those in a hierarchical manner, where one decision boundary is learnt at a time, could lead to improved efficacy. In particular, we propose a two-stage hierarchical framework consisting only of binary classifiers, where the first stage learns the boundary between low and high quality, and the second stage learns the boundary differentiating between the extremes within the high (as well as low) quality subrange. The boundary obtained in the first stage is eventually ignored, and the two boundaries obtained in the second stage are taken as the desired ones of the usable category. In this paper, we seek to establish the hypothesis of superiority of the proposed scheme.

With the recent advent of explainable DL [10, 13], one now seeks a statistically accurate DQA tool that also provides a clinically meaningful visual explanation behind each quality decision. Such explanations not only assist human operator in authenticating algorithm-assigned quality labels, but also help in debugging and developing generalizable DQA algorithms that are data-agnostic, clinically reliable and practically deployable. Research into explainable DL models for DQA of FP images has begun only in the last couple of years. In one work [48], possible usefulness of machine-generated explanations has been illustrated using gradient-based class activation maps (Grad-CAM) [40]. In this paper, we also adopt Grad-CAM explanations, and provide overlayed visualization using heat and transparency maps. Further, we attempt at relating such explanations to the DQ grading criteria and considerations, and hence studying their clinical relevance. Encouragingly, our explanations corroborate expert intuitions in cases of not only the extreme quality labels of good and unusable but the intermediate label of usable as well.

## 2. Materials and Methods

The proposed workflow consisting of a two-stage hierarchical DL classifier and followed by Grad-CAM-generated

**Table 1**
Composition of EyeQ Dataset [11]

|       | Good  | Usable | Unusable | Total |
|-------|-------|--------|----------|-------|
| **Train** | 8347  | 1896   | 2320     | 12543 |
| **Test**  | 8471  | 4558   | 3220     | 16249 |
| Total | 16818 | 6434   | 5540     | 28792 |

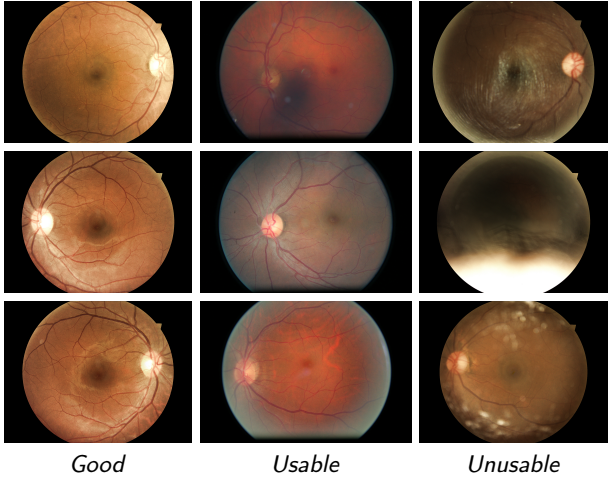

*Good*  *Usable*  *Unusable*

**Figure 5:** Illustrative examples of FPs graded by expert.

visual explanations is schematically presented in Fig. 4, and subsequently elaborated. First we describe the EyeQ dataset, on which the proposed DL classifier was trained and tested.

### 2.1. EyeQ Dataset

As alluded earlier, EyeQ dataset consists of 28,792 FP images divided into three DQ categories, viz. good, usable, and unusable ('reject') [11]. Respectively, 12543 and 16249 images are earmarked as training and test sets with counts of various subdivisions given in Table 1. Grading was performed by taking into account the presence of indicators (viz. imaging artifacts, blur, contrast and illumination) of poor quality as well as the visibility of ocular structures (viz. OD, macula and retinal vessels) as follows.

- *Good:* no low-quality indicators and clear visibility of all ocular structures (first column in Fig. 5).

- *Usable:* presence of all ocular structures, and low-quality indicators affecting 20% of the fundus region or less (second column in Fig. 5).

- *Unusuable (reject):* Missing optic disc and/or macula, or significant presence (more than 20%) of low-quality indicators (third column in Fig. 5).

### 2.2. Building blocks of hierarchical DL Classifier

The proposed two-stage hierarchical classifier, depicted in Fig. 4, consists of three binary classifier models, viz. Model-1 (first stage), Model-2 and Model-3 (second stage), which distinguish, respectively, between high and low DQ, between good and usable DQ within the high DQ category, and between usable and unusable DQ within the low DQ category.

*Proposed approach:* We investigated the end-to-end efficacy of the proposed classifier considering various versions, where each of the aforementioned models made use of one of the two legacy DL architectures, DenseNet (dense convolutional network) and EfficientNet [16, 45], with suitable hyperparameters and appropriate modifications. In general, design of a DL model involves optimal choice of the three dimensions, viz. input image resolution, the number of layers (depth), and the number of filters in each layer (width). The depth and the width determine the respective ability to learn the rich and complex features, and fine-grained features within the limit set by the input image resolution [45]. Most legacy networks have been developed for specific broad tasks, and input resolution, depth and width are chosen via trials and errors based on large standardized datasets. Note that such a legacy network with pretrained parameters, including filter coefficients and model weights, is usually adopted as the basis for transfer learning, where additional fully connected (FC) layers are appended and trained to tune to the specific task at hand. A legacy network, the DenseNet, uses skip connections, and handles large depths, although the benefit tapers off beyond certain limits. In contrast, the EfficientNet scales the three aforesaid dimensions in a compound manner according to resource constraints. Both networks, originally trained on the ImageNet database of natural images [7, 20], have recently been repurposed for use in DQA of FP images [48, 29]. While the variant DenseNet121 was the original recommendation for use on the EyeQ dataset [11], the base variant EfficientNet B0 has achieved high performance on a private FP DQ dataset [29].

*DenseNet121:* With increased number of layers in a DL network, the amount of information about the input (resp. gradients while back propagation) may vanish as one approaches the final (resp. the beginning) layers, posing difficulty in the training process. Since the last few years, residual networks with skip connections in the feed forward path have proven to alleviate such difficulty [14]. Further improvement has recently been achieved by DenseNet [16], schematically depicted in Fig. 6(a), where layer $L_n$ in a dense block takes as input a composite function of the concatenation of feature maps of all preceding layers $L_{n-1}, L_{n-2}, ..., L_1$ via skip connections. The said function sequentially performs batch normalisation, rectified linear unit (ReLU) function ($\max(0, x)$ in variable $x$) and $3 \times 3$ convolution. As mentioned earlier, we considered the variant DenseNet121, which takes one $224 \times 224$ 3-channel color image at a time as input, and has four dense blocks, and a total of 121 layers and $6.96M$ parameters [16, 48].

*EfficientNet B0:* The EfficientNet employs compound scaling of the three aforesaid dimensions, and caters to practical resource constraints, while maintaining model efficiency [45]. The original variant EfficientNet B0 considers a baseline architecture MobileNetV1 [15], and performs compound scaling to optimize the three dimensions to meet the computational resource constraint (Fig. 6(b)). Subsequent variants B1, ..., B7 are increasingly complex, and are not considered in this paper. Our choice, the variant B0, takes as input 224×224 3-channel color images as earlier, and has a total of 237 layers, and $5.3M$ parameters [45]. Overall, EfficientNet B0 is smaller than DenseNet121.

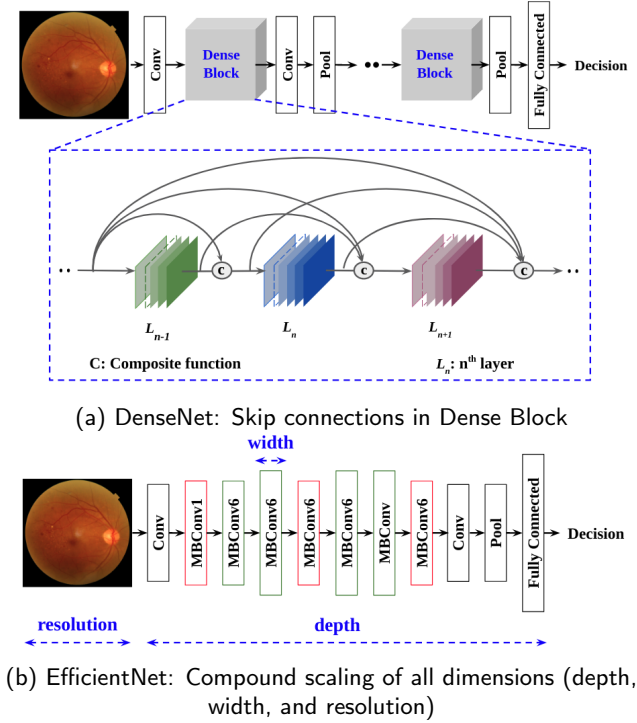

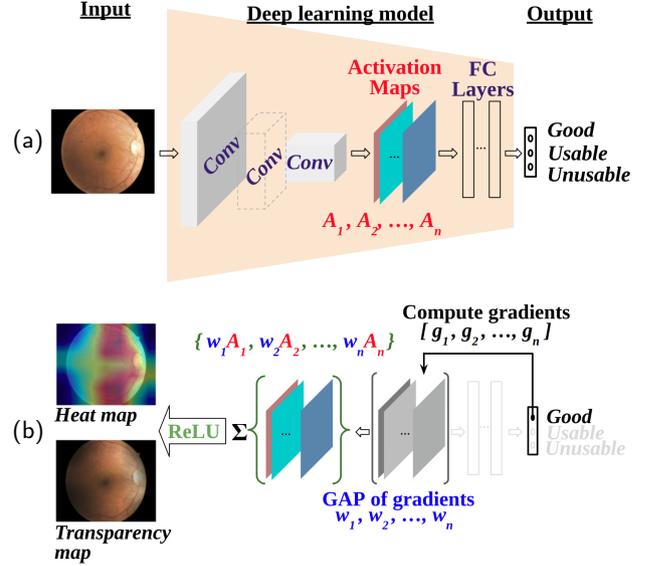

**Figure 6:** Block schematic of legacy deep learning models: (a) DenseNet and (b) EfficientNet.

**Figure 7:** Schematic pipeline of (a) DL-based prediction, and (b) Grad-CAM-generated explanation for predicted quality label (illustration for predicted label good) along with overlayed heat and transparency maps.

### 2.3. Evaluation Methods

As alluded earlier, we considered the proposed two-stage hierarchical classifier with each component model being either DenseNet121 or EfficientNet B0 appended with additional FC layers for transfer learning. Specifically, we made use of the EyeQ dataset, and trained appended model parameters using the earmarked training set (see Table 1). The performance of the trained hierarchical classifier was evaluated on the test set as described next.

*Performance measures:* We carried out comparative performance analysis of the proposed 3-class classifier in terms of the 3×3 confusion matrix $C$ (the $(i, j)$-th element $C_{ij}$ indicating the number of images of class $i$ predicted to belong to class $j$), and its derivatives accuracy, precision, recall and $F$-Score [44]. More precisely, we consider normalized confusion matrix $\bar{C}$, defined by the $(i, j)$-th element $\bar{C}_{ij} = C_{ij}/\sum_{k=1}^{3} C_{ik}$ (indicating the probability of predicting an image of class $i$ as belonging to class $j$). Clearly, each row of $\bar{C}$ always sums to one, and we desire each diagonal entry ($\bar{C}_{ii}$, $i = 1, 2, 3$) to be close to its ideal value one. Further, accuracy is defined by the fraction of times the classifier correctly predicts the class of an image, i.e., $\sum_{i=1}^{3} C_{ii}/\sum_{i=1}^{3}\sum_{j=1}^{3} C_{ij}$, and equals one for a perfect classifier. Further, precision and recall are averaged class-conditional measures, defined respectively by the average ratio of true predictions to all predictions of specific class, i.e., $\frac{1}{3}\sum_{j=1}^{3}(C_{jj}/\sum_{l=1}^{3} C_{lj})$, and the average class-conditional accuracy, i.e., $\frac{1}{3}\sum_{i=1}^{3}(C_{ii}/\sum_{k=1}^{3} C_{ik}) = \frac{1}{3}\sum_{i=1}^{3} \bar{C}_{ii}$. Precision and recall, both with ideal values equal to one, trade off against each other in case of practical classifiers. Finally, $F$-Score, defined by the harmonic mean of precision and recall, provides a standard measure of overall performance.

*Visual explanations:* At their inception, inner workings of high-performance DL models, owing to their complex architecture (consisting of convolution blocks, activation maps, FC layers and other components as shown in Fig. 7(a)), were not amenable to human intuition, and the outcomes could not be authenticated [3]. Machine-generated explanations, such as those generated by the ubiquitous gradient-based Grad-CAM technique [40], have begun to overcome the aforementioned limitation. Specifically, as shown in Fig. 7(b), gradients $g_1, g_2, ..., g_n$ ($n$ being the number of activation maps) corresponding to a specific class ('good', in the illustration) were computed with respect to respective activation maps $A_1, A_2, ..., A_n$ of the final convolutional layer. Importance weights $w_1, w_2, ..., w_n$ corresponding to the activation maps are obtained via global average pooling (GAP) of the gradients. Subsequently, the weighted sum $\sum_k w_k A_k$ was computed and passed through the ReLU function ($\max(0, x)$ in variable $x$) to take only positive correlations into account. The resulting map is finally up-sampled to the input image size, and overlaid on the input image as a heat-map of explanation, with 'hotter' shades indicating higher relevance. Although Grad-CAM-based heat-maps have recently been used for DQA of FP images [48],

**Table 2**

Performance comparison with other methods on EyeQ Dataset (superior values are bold-faced and second best values are italicized)

|  | Algorithm | Accuracy | Precision | Recall | $F$-Score |
|---|---|---|---|---|---|
| *NIQA* | BRISQUE, Mittal et al. (2012)* [25] | 76.92 | 76.08 | 70.95 | 71.12 |
|  | NBIQA, Ou et al. (2019)* [32] | 79.17 | 76.41 | 75.09 | 74.41 |
|  | TS-CNN, Yan et al. (2018)* [49] | 79.26 | 79.76 | 74.46 | 74.81 |
| *FP DQA* | HVS-based, Wang et al. (2015)* [47] | – | 74.04 | 69.45 | 69.91 |
| *(Reported work)* | ScatNet, Dev et al. (2019)** [8] | 77.93 | 76.97 | 72.27 | 74.54 |
|  | MCS-augmented, Manne et al. (2021)** [23] | 84.23 | 82.80 | 80.20 | 81.48 |
|  | MCF-Net, Huazhu et al. (2019)† [11] | 88.00 | 86.51 | 85.74 | 86.12 |
|  | MR-CNN, Aditya et al. (2020) [37] | 88.43 | 86.97 | 87.00 | 86.94 |
|  | DB-SalStructIQA, Xu et al. (2020) [48] | *88.97* | 87.48 | 87.21 | *87.23* |
| *(Present work)* | Each component model being DenseNet121 | 88.90 | *87.71* | 86.73 | 87.22 |
|  | Each component model being EfficientNet B0 | 88.93 | 86.97 | *87.35* | 87.16 |
|  | **Recommended combination**‡ | **89.44** | **87.98** | **87.70** | **87.84** |
|  | (% gain over the present state of the art [48]) | (0.53) | (0.57) | (0.56) | (0.70) |

* Results reproduced from that of reported in [11, 48]
** Results reproduced by authors on EyeQ dataset using reported feature set
† Updated results from https://github.com/HzFu/EyeQ
‡ Model-1: DenseNet121; Model-2:DenseNet121; Model-3: EfficientNet B0.

such explanations have not been correlated with various DQ grading criteria (such as visibility of occular structures in case of good DQ images, and presence of artifacts or uneven illumination and contrast in case of unusable DQ images). As an improvement, we generate Grad-CAM explanations based on the proposed hierarchical architecture, and correlate those systematically with the aforesaid grading criteria. Specifically, note that each of our explanations pertains to one of the two classifiers, Model-2 and Model-3, in the second stage (see Fig.4), distinguishing between good and usable, and usable and unusable images, respectively, and hence conveys specific information within a relevant sub-range (either 'high' or 'low' quality) of the DQ continuum. We observe that such specific explanations are close to the subjective experience of DQ. We also provided visualization of the said explanations as overlaid transparency maps that allow greater visibility of features with higher relevance to the DQ category prediction. Such visualization, extending the natural way of viewing FP images, could be attractive to medical personnel.

## 3. Experimental Results

In this section we present the experimental results and compare the performance of the proposed hierarchical DL classifier on EyeQ dataset. As mentioned earlier, each of the components, viz. Model-1, Model-2 and Model-3, uses as the base network either DenseNet121 or EfficientNet B0.

*Statistical performance comparison:* In Table 2, we present the performance of combinations of such legacy networks in comparison with reported performance of existing techniques. Competing techniques include NIQA models such as blind/referenceless image spatial quality evaluator (BRISQUE) [25], novel blind IQA (NBIQA) [32], and two-stream convolutional networks (TS-CNN) [49]. Performance comparison is also made with FP DQA tools taking classical approaches based on human visual system (HVS) based hand-crafted features [47], structure preserving scattering network (ScatNet) features [8] and multi-color space (MCS) augmented features [23], as well as DL approaches including MCS fusion network (MCF-Net) [11], multivariate regression-based convolutional neural network (MR-CNN) [37], and dual-branch salient structure IQA (DB-SalStructIQA) [48].

Compared to FP DQA algorithms, the NIQA algorithms generally exhibited inferior performance in Table 2 in terms of various measures, as anticipated. Further, among FP DQA algorithms, the existing ones employing DL [11, 37, 48] outperformed those using handcrafted features [47, 8, 23]. The proposed hierarchical DL classifier using only either DenseNet121 or EfficientNet B0 in all of the component models performed close to the current state of the art [48]. However, choosing the optimal combination for the component models, an improvement ranging between 0.53% and 0.70% was obtained over the state of the art in terms of accuracy, precision, recall and $F$-score.

*Visual explanations:* We made a subjective assessment of the Grad-CAM-based explanations generated for different DQ categories, and depict in Fig. 8 overlaid heat and transparency maps on representative FP images. Visual explanations of images, predicted to be of good DQ, with examples shown in Figs. 8(a) and 8(b), are generated by Model-2, and tend to highlight regions with clearly visible ocular structures such as OD, macula and retinal vessels. On the other hand, those of images, predicted to be unusable, with examples shown in Figs. 8(e) and 8(f), are generated by Model-3, and tend to highlight regions with relevant low-quality indicators such as improper contrast and illumination. In contrast to the preceding, recall from Fig. 4 that images with usable DQ could be predicted either by Model-2 (with an example shown in Fig. 8(c)) or by Model-3 (with an example shown in Fig. 8(d)), and the corresponding visual explanations need to be interpreted differently. In the former, why an image did not have high DQ was explained by highlighting regions of low-quality indicators, while in the latter, why an image is still usable was explained by highlighting regions of clearly visible ocular structures. Overall, machine-generated explanations across DQ categories provided visual DQ cues, which would potentially assist medical decision making.

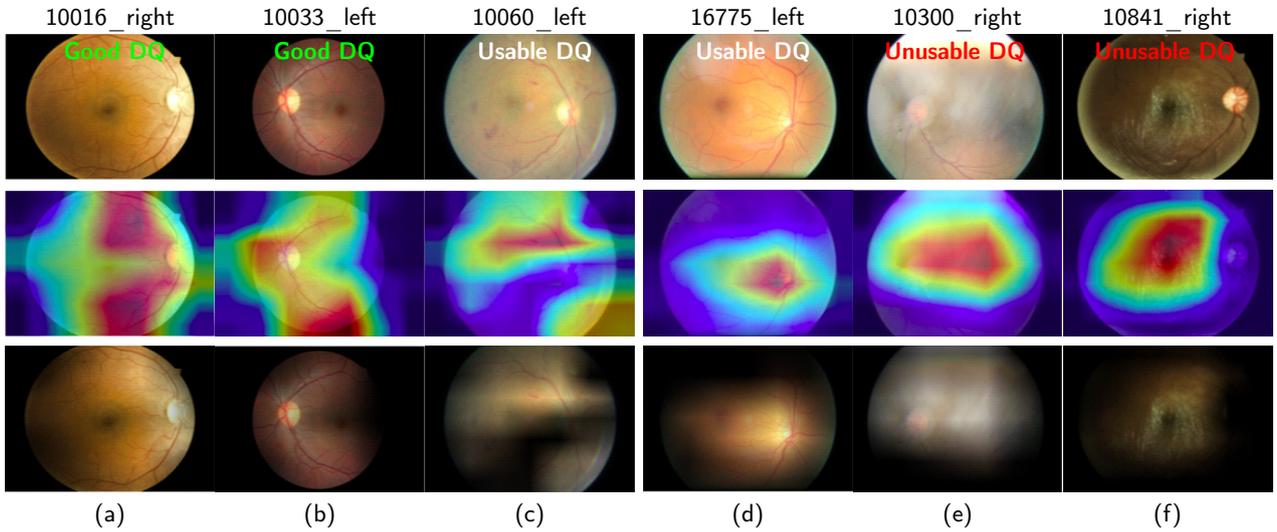

**Figure 8:** Representative FP images of three DQ categories with corresponding Grad-CAM explanations. Columns: (a), (b) belong to Good DQ and corresponding explanations were generated using Model-2; (c) and (d) belong to Usable DQ with corresponding explanations obtained using Model-2 and Model-3, respectively; (e), (f) belong to Good DQ and corresponding explanations were generated using Model-3. Rows: first row correspond to actual FP images, second and third row, respectively represent the heatmap and transparency map, corresponding to the actual image presented in the first row.

## 4. Discussion and Future Course

In this paper, we proposed a hierarchical DL architecture for DQA of FP images into three DQ classes, viz. good, usable and unusable. Specifically, we demonstrated that a suitable combination of DenseNet121 and EfficientNet B0 exhibited superior performance in terms of various measures related to the confusion matrix. Accompanying the predicted DQ label of an FP image, we generated Grad-CAM-based visual explanations that assume importance by significantly correlating with expert intuitions. Although we currently presented qualitative correlation, we plan to explore quantitative criteria in the future.

Beyond overall performance measures, class-specific information could also be useful in performance comparison, when the cost of making an error differs among classes. Indeed, consider the following two scenarios of teleophthalmology, where the availability of (i) the expert and (ii) the machine are limited. In each, the cost of misclassifying the DQ label depends on the actual class, and the class-conditional costs depend on the scenarios as follows. In the first scenario, one should accurately identify unusable images, and discard those so as to spare the expert avoidable burden. In the second scenario, contrarily, one aims at minimizing machine usage by avoiding usable images being incorrectly labeled unusable (and hence recapture) as much as possible. A given accuracy level (e.g., 88.76% as well as 89.03%) is achieved by a range of confusion matrices, among which, as given in Table 3, $C_1$ (resp. $C_2$) maximizes the class-conditional accuracy for the unusable (resp. usable) class, and hence suits the first (resp. second) scenario. Our analysis could be attractive due to the flexibility of catering to such class-specific considerations.

We envisage a future with FP cameras equipped with resident DQA algorithms similar to proposed one. Such smart cameras would not only enhance the efficacy of a teleophthalmology system but also assist optometrists located at primary, secondary and tertiary centers, by providing accurate feedback on DQ, and hence obviating acquisition of unnecessary images. Further, at organisations such as the University of Pittsburgh Medical Center (UPMC) that acquire thousands of FP images daily, such automated tool would potentially help in distinguishing good DQ images for conducting large-scale retrospective studies. Further, studies on DQA for FP images have so far been conducted in a generic view-independent framework. However, diagnosis of specific diseases demand view-specific (such as macula centered, OD centered, and so on) FP images. In future, we seek to extend the present analysis, and develop view/disease-specific DQA tools.

**Table 3**
Various confusion matrices for fixed level of accuracy (%) [Maximised class-conditional values are boldfaced]

|  | $Acc = 88.76 \pm 0.03$ | | | $Acc = 89.03 \pm 0.03$ | | |
|---|---|---|---|---|---|---|
|  | *Good* | *Usable* | *Unusable* | *Good* | *Usable* | *Unusable* |
| $C_1$ | 94.73 | 5.19 | 0.07 | 94.70 | 5.23 | 0.07 |
|  | 10.93 | 75.80 | 13.27 | 10.68 | 77.56 | 11.76 |
|  | 0.62 | 8.11 | **91.27** | 0.62 | 9.10 | **90.28** |
| $C_2$ | 92.41 | 7.54 | 0.05 | 93.38 | 6.47 | 0.15 |
|  | 8.89 | **83.41** | 7.7 | 9.79 | **82.16** | 8.05 |
|  | 0.56 | 12.58 | 86.86 | 0.40 | 12.40 | 87.20 |

## 5. Broader Impact on Clinical Practice

With increasing awareness of retinal diseases and upcoming treatment options, there has been an overall increase in the burden on eye hospitals. Remote testing and teleophthalmology are being adapted in retina clinics, particularly during and after the pandemic. Prior to the pandemic, telecommunication in ophthalmology commonly took place through the "Store-and-Forward" technique, where information (e.g., retinal image) was sent and reviewed later [39]. As in-person clinic visits became limited, synchronous, bi-directional teleophthalmology appointments became more common and implementation of tele-screening allowed for effective triaging for in-person visits [39, 5]. Retina emergencies such as retinal detachment often require emergent retina specialist evaluation and surgery to prevent permanent vision loss. The implementation of tele-screening for ocular emergencies as seen during the COVID-19 outbreak appears to be an effective method for reducing the time to care, thus, seemingly advantageous for retinal emergencies requiring immediate intervention. Thus, an immediate feedback, as proposed in this paper, to the operator about diagnostic quality helps one obtain "usable" images and serves the purpose of tele-screening.

## 6. Acknowledgements

The work was partly supported by Grant BT/PR16582/BID/7/667/2016, Department of Biotechnology (DBT), Ministry of Science and Technology, the Government of India. Manne thanks the Ministry of Electronics and Information Technology (MeitY), the Government of India, for fellowship grant under Visvesvaraya PhD Scheme for Electronics and IT. Manne also thanks Abhishar Sinha and Surya Koidala for their inputs in this project.